\documentclass[aps,prb,floats,epsf,twocolumn]{revtex4}
\usepackage{graphicx,epsf}
\usepackage{amsfonts}
\begin{document}

\title{Theory of interacting electrons on the honeycomb lattice}

\author{Igor F. Herbut, Vladimir Juri\v ci\' c, and Bitan Roy}

\affiliation{Department of Physics, Simon Fraser University,
 Burnaby, British Columbia, Canada V5A 1S6}

\begin{abstract}
The general low-energy theory of electrons interacting via repulsive short-range interactions on graphene's honeycomb lattice at half filling is presented. The exact symmetry of the Lagrangian with local quartic terms for the Dirac four-component  field dictated by the lattice is identified as $D_2\times U_c (1)\times$time reversal, where $D_2$ is the dihedral group, and $U_c(1)$ is a subgroup of the $SU_c (2)$ "chiral" group of the non-interacting Lagrangian, that represents translations in Dirac language. The Lagrangian describing spinless particles respecting this symmetry is parameterized  by six independent coupling constants. We show how first imposing the rotational, then Lorentz, and finally chiral symmetry to the quartic terms - in conjunction with the Fierz transformations - eventually reduces the set of couplings to just two, in the "maximally symmetric" local interacting theory. We identify the two critical points in such a Lorentz and chirally symmetric theory as describing metal-insulator transitions into the states with either time-reversal or chiral symmetry being broken.  The latter is proposed to govern the continuous transition in both the Thirring and Nambu-Jona-Lasinio models in 2+1 dimensions and with a single Dirac field. In the site-localized, "atomic", limit of the interacting Hamiltonian, under the assumption of emergent Lorentz invariance, the low-energy theory describes the continuous transitions into the insulator with either a finite Haldane's (circulating currents) or Semenoff's  (staggered density) masses, both in the universality class of the Gross-Neveu model. The simple picture of the metal-insulator transition on a honeycomb lattice emerges at which the residue of the quasiparticle pole at the metallic, and the mass-gap in the insulating phase both vanish continuously as the critical point is approached. In contrast to these two critical quantities, we argue that the Fermi velocity is non-critical as a consequence of the dynamical exponent being fixed to unity by the emergent Lorentz invariance near criticality. Possible effects of the long-range Coulomb interaction, and the critical behavior of the specific heat and conductivity are discussed.
\end{abstract}
\date{\today}
\maketitle

\vspace{10pt}

\section{Introduction}

Two-dimensional honeycomb lattice of carbon atoms may be viewed as the mother of all other forms of carbon. Its crucial electronic property, which arises as a consequence of the absence of the inversion symmetry around the lattice site, is that the usual Fermi surface is reduced to just two points. The electronic dispersion may be linearized around these two points, after which it becomes isotropic and dependent on the single dimensionful parameter, Fermi velocity $v_F \approx c/300$. The pseudo-relativistic nature of the electronic motion in graphene has since its synthesis placed this material at the center stage of condensed matter physics. Many qualitatively novel phenomena that take, or may take place in such a system of "Dirac" electrons are actively  discussed in the rapidly growing literature on the subject. \cite{rmp}

   In this paper we discuss the low-energy theory and the metal-insulator quantum phase transitions of the {\it interacting} Dirac electrons on the honeycomb lattice, building upon and expanding significantly the earlier work by one of us. \cite{herbut1} In the first approximation, {\it all} weak interactions of Dirac electrons in graphene may be neglected at half filling, when the Fermi surface consists of the Dirac points. This is because short-range interactions are represented by local terms which are quartic in the electron fields, which makes them irrelevant near the non-interacting fixed point by power counting. The same conclusion turns out to apply to the long-range tail of the Coulomb interaction, which remains unscreened in graphene, although only marginally so. \cite{gonzales, herbut1, vafek} Nevertheless, if strong enough, the same interactions would turn graphene into a gapped Mott insulator. As an example, at a strong on-site repulsion the system is likely to be the usual N\'{e}el antiferromagnet. \cite{herbut1, herbutAF} It is not {\it a priori} clear on which side of this metal-insulator transition graphene should be. With the standard estimate for the nearest-neighbor hopping in graphene of $t=2.5 eV$ and the Hubbard interaction of $U\approx 7-12 eV$, it seems that the system is below yet not too far from the critical point estimated to be at $U/t \approx 4 - 5$.  \cite{herbut1,sorella,martelo, paiva} If sufficiently weak, the electron-electron interactions only provide corrections to scaling of various quantities, which ultimately vanish at low temperatures or frequencies.  At, what is probably  a more realistic, an intermediate strength, the flow of interactions and the concomitant low-energy behavior may be influenced by the existence of metal-insulator critical points. It is possible that some of the consequences of such interaction-dominated physics have already been observed in the quantization of the Hall conductance at filling factors zero and one. \cite{zhang1, ong, khveshchenko, gusynin, herbut2, alicea}  As we argued elsewhere, the anomalously large value of the minimal conductivity in graphene \cite{geim} may be yet another consequence of the Coulomb repulsion between electrons. \cite{herbut4,sachdev}

The above discussion raises some basic questions. What is the minimal description of interacting electrons in graphene at "low" energies? What is the symmetry of the continuum interacting theory, and how does it constrain the number of  coupling constants?  What kinds of order may be expected at strong coupling, and what is the nature of the metal-insulator quantum phase transition? In this paper we address these and related issues. In the rest of the introduction we give a preview of our main results.

The simplest prototypical system that exhibits the physics of interacting Dirac fermions which we seek to understand is the collection of spinless electrons interacting via short-range interactions, at half-filling. For present purposes an interaction may be considered as  "short-ranged" if its Fourier transform at the vanishing wavevector is finite. \cite{remark1} The least irrelevant quartic terms one can add to the non-interacting Dirac Lagrangian will then be local in space-time, and of course quartic in terms of the four-component Dirac fields that describe the electronic modes near the two inequivalent Dirac points at wavevectors $ \pm \vec{K}$ at the edges of the Brillouin zone. The most general local quartic term in the Lagrangian would be of the form
\begin{equation}
L_{int} = ( \Psi^\dagger (\vec{x},\tau) M_1 \Psi(\vec{x},\tau) ) ( \Psi^\dagger(\vec{x},\tau) M_2 \Psi(\vec{x},\tau) ),
\end{equation}
where $M_1$ and $M_2$ are four-dimensional Hermitian matrices.  The symmetry alone, however, immediately drastically reduces the number of independent couplings from the apparent $136$ to just fifteen. Although the point group of the honeycomb lattice is $C_{6v}$, the exact spatial discrete symmetry of the Lagrangian is only the {\it dihedral group}  $D_2$, or the {\it vierergruppe}, which consists of the reflections through the two coordinate axis shown in Fig. 1, and the inversion through the origin. Such a small symmetry results from the very choice of two inequivalent Dirac points out of six corners of the Brillouin zone, which reduces the symmetry to the simple exchange of the two sublattices (reflection around A axis), the exchange of Dirac points (reflection around B axis), and their product (the inversion through the origin). $D_2$, the time-reversal, and the translational invariance are shown to leave fifteen possible different local quartic terms in the Lagrangian.

  \begin{figure}[t]
{\centering\resizebox*{85mm}{!}{\includegraphics{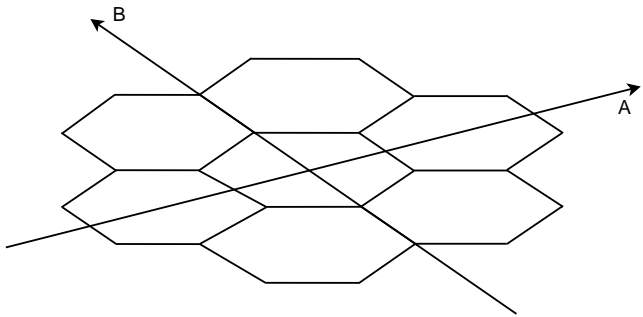}}
\par} \caption[] {Two axis of symmetry of the low-energy theory of graphene, in real space. The Dirac points in this coordinate frame are at $\pm \vec{K} = (1,0) (4\pi/3 a)$, i. e. along the A-axis.}
\end{figure}

   Fortunately, not all of these still numerous quartic terms are independent, and there are linear constraints between them implied by the algebraic Fierz identities. \cite{takahashi} The Fierz transformations are rewritings of a given quartic term in terms of others, and we provide the general formalism for determining the number and the type of independent quartic couplings of a given symmetry. For the case at hand we find that spinless electrons interacting with short-range interactions on honeycomb lattice are in fact described by only {\it six} independent local quartic terms. The inclusion of electron spin would double this number to twelve.

   The linearized noninteracting Lagrangian for Dirac electrons,
   \begin{equation}
   L_0 = \bar{\Psi} (\vec{x},\tau) \gamma_\mu \partial_\mu \Psi (\vec{x},\tau)
   \end{equation}
   as well-known, exhibits the Lorentz and the global $SU_c (2)$ ("chiral") symmetry. The latter, generated   by $\{ \gamma_3, \gamma_5, \gamma_{35} \}$, with $\gamma_{35}= -i\gamma_3 \gamma_5$, is nothing but the "rotation" of the "pseudospin", or "valley", corresponding to two inequivalent Dirac points. \cite{dwave} A general quartic term allowed by the lattice symmetry, on the other hand, has a much smaller symmetry, as already mentioned. Nevertheless, we will argue that near the metal-insulator quantum critical points, all, or nearly all of the larger symmetry possessed by the non-interacting part of the Lagrangian gets restored. This conclusion is supported by the, admittedly uncontrolled, but nevertheless quite informative one-loop calculation. First, we find three distinct critical points in the theory, all of which have not only the rotational, but the full Lorentz-symmetric form. This immediately implies that the dynamical critical exponent is always $z=1$. This is quite remarkable in light of the fact that the microscopic theory is not even rotationally invariant, and that the critical points in question are purely short-ranged. \cite{remark2} The fact that $z=1$ has important implications for several key physical observables near the critical point, as we discuss shortly. Furthermore, we find that two out of three critical points in the theory exhibit a full chiral symmetry as well. We identify the three fixed points  in the theory as corresponding to three possible order parameters, or "masses" that develop in the insulating phase at strong coupling.

   1) $\langle \bar{\Psi} \gamma_{35} \Psi \rangle $, which preserves chiral, but breaks time-reversal symmetry. Microscopically, this order parameter may be understood as a specific pattern of circulating currents, as discussed in the past. \cite{haldane}

   2) $\langle \bar{\Psi} \Psi \rangle $, which preserves the time-reversal symmetry, and the single chiral generator $\gamma_{35}$, which will be shown to correspond to translational invariance. This order parameter describes a finite staggered density, i. e. the difference between the average densities on the two sublattices of the honeycomb lattice. \cite{semenoff}

   3) $\langle \bar{\Psi} (\gamma_3 \cos\alpha + \gamma_5 \sin\alpha) \Psi \rangle $, which preserves the time-reversal, but breaks translational invariance ($\gamma_{35}$). This order parameter can be understood as the  specific "Kekule" modulation of the nearest-neighbor hopping integrals. \cite{hou}

    In one-loop calculation all three critical points have the same correlation length exponent $\nu=1$, which we believe is an artifact of the quadratic approximation. The result that the dynamical critical exponent $z=1$ is, on the other hand, possibly exact. If we denote the relevant interaction parameter with $V$, the Fermi velocity near the transition scales as
   \begin{equation}
   v_F \sim (V_c - V)^{\nu (z-1)}
   \end{equation}
   so the above value of $z$ would simply imply that it stays regular at the transition. This appears to be in agreement with the picture of the transition as the opening of the relativistic "mass" in the spectrum. The mass-gap in the insulating phase scales as usual \cite{book} as
   \begin{equation}
   m\sim (V-V_c)^{z\nu}.
   \end{equation}
The transition on the metallic side is manifested as vanishing of the residue of the quasiparticle pole \cite{brinkman}
\begin{equation}
Z\sim (V_c -V)^{\nu \eta_\Psi}.
\end{equation}
where we assumed $z=1$. (A more general power-law is discussed in the text.) At one-loop the fermion anomalous dimension $\eta_\Psi$ vanishes, but in general it is a positive, small, and critical-point-dependent number. The overall picture of the metal-insulator transition that emerges is presented in Fig. 2.

   \begin{figure}[t]
{\centering\resizebox*{85mm}{!}{\includegraphics{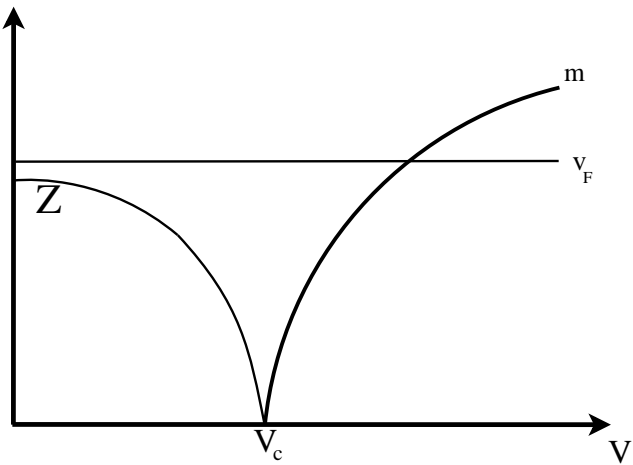}}
\par} \caption[] {The behavior of the the Fermi velocity($v_F$), strength of the quasiparticle pole ($Z$), and the gap ($m$) near the metal-insulator transition.}
\end{figure}

 For graphene's $p_z$-orbitals well localized on carbon sites, a further significant simplification takes place. All the terms without the equal number of creation and annihilation operators for each of the two sublattices must vanish. Assuming again the emergent Lorentz symmetry at low energies this allows one to finally write the simplest internally consistent interacting theory as
 \begin{equation}
 L= L_0 +  g_{D2} (\bar{\Psi}\gamma_{35} \Psi)^2 + g_{C1} (\bar{\Psi} \Psi)^2.
 \end{equation}
 This Lagrangian provides the minimal low-energy description of interacting spinless electrons on honeycomb lattice. It has two critical points, corresponding to transitions into insulators 1) and 2) in the above, both corresponding to the Gross-Neveu criticality in 2+1 dimensions. We discuss the internal consistency and the sufficiency of this Lagrangian and some of the peculiarities of the ensuing phase diagram.

   The rest of the paper is organized as follows. We discuss the point symmetry, translational symmetry, and the time-reversal symmetry of the interacting Lagrangian as dictated by the microscopic Hamiltonian for the system in the next section. In section III it is shown how further enlargements of the symmetry would reduce the number of coupling constants. We introduce the notion of "maximally symmetric" theory, which shares the full Lorentz and chiral symmetry with the quadratic term in the Lagrangian. The general formalism of Fierz transformations is developed and applied to the cases of interest in section IV. The change of the coupling constants with the ultraviolet cutoff in the theory is  studied in section V. The atomic limit of the general interacting theory is described in section VI, and the critical exponents are discussed in section VII. In section VIII we discuss the scaling of the electron propagator and the power-laws for various quantities of interest. The  discussion of the long-range Coulomb interaction and the critical behavior of the specific heat and the optical conductivity are given in section IX, and the summary in section X. Finally, in Appendixes we present some of the requisite technical details: the Fierz transformation, the spectral form of the asymmetric matrix needed in section IV, and an alternative implementation of the renormalization group in presence of linear constraints.

\section{Symmetries and short-range interactions}

\subsection{Hamiltonian and the Lagrangian}

As the simplest microscopic model that contains the relevant physics we may
consider the tight-binding Hamiltonian on the graphene's honeycomb lattice, defined as
\begin{equation}\label{Ht}
H_0= \tilde{t} {\sum_{\vec{A},i}}{u^\dagger}
(\vec{A})v(\vec{A}+\vec{b_i})+H.c.,
\end{equation}
where $u$ and $v$ are the electron annihilation operators at two
triangular sublattices of the honeycomb lattice. Here,
$\vec{A}$ denotes sites of the sublattice generated by linear
combinations of basis vectors $\vec{a_1}=(\sqrt{3},-1)a$,
$\vec{a_2}=(0,1)a$, whereas $\vec{B}=\vec{A}+\vec{b}$ are the
sites on the second sublattice, with $\vec{b}$ being
$\vec{b_1}=(1/\sqrt{3},1)a/2, \vec{b_2}=(1/\sqrt{3},-1)a/2$, or
$\vec{b_3}=(-1/\sqrt{3},0)a$, and $a$ is the lattice spacing.

Within the framework of the tight-binding model the energy
spectrum is doubly degenerate
$E(\vec{k})=\pm\tilde{t}|{\sum_i}\exp[\vec{k}\cdot\vec{b_i}]|$, and
becomes linear and isotropic in the vicinity of six Dirac points,
at the edge of the Brillouin zone, among which only two, hereafter chosen to be at
$\pm{\vec{K}}$ with $\vec{K}=(1,1/\sqrt{3})(2{\pi}/a\sqrt{3})$,
are inequivalent. Retaining only the Fourier components in the
vicinity of these two inequivalent points, the quantum
mechanical action corresponding to $H_0$ at low energies can be
written in the form $S={\int_0}^{1/T}{d\tau}{d\vec{x}}L_0$, with
the free Lagrangian density $L_0$ defined as in Eq. (2), with
 $\tau$ as the imaginary time and $T$ is the temperature. Matrices
$\gamma_\mu$ satisfy the Clifford algebra
$\{{\gamma_\mu},{\gamma_\nu}\}=2{\delta_{{\mu}{\nu}}}$,
${\mu,\nu}=0,1,2$, $\bar{\Psi}=\Psi^\dagger \gamma_0$. The summation over
repeated space-time indices is assumed hereafter. The fermionic field
$\Psi({\vec x},\tau)$ is defined as
\begin{widetext}
\begin{equation}\label{Dirac-spinor}
\Psi^\dagger(\vec{x},\tau)=T{\sum_{\omega_n}}{\int^\Lambda}{\frac{d\vec{q}}{(2\pi
a)^2}}e^{{i{\omega_n}\tau}+i{\vec{q}\cdot{\vec{x}}}}(u^\dagger(\vec{K}+\vec{q},\omega_n),
v^\dagger(\vec{K}+\vec{q},\omega_n),
u^\dagger(-\vec{K}+\vec{q},\omega_n),
v^\dagger(-\vec{K}+\vec{q},\omega_n)).
\end{equation}
\end{widetext}
Here, the reference frame is conveniently rotated so that
$q_x=\vec{q}\cdot\vec{K}/K$,
$q_y=(\vec{K}\times\vec{q})\times{\vec{K}}/K^2$,
$\omega_n=(2n+1)\pi T$ are the fermionic Matsubara frequencies,
$\Lambda\sim 1/a$ is a high-energy cutoff, and we set $\hbar=k_B=v_F=1$,
where $v_F=\tilde{t}a\sqrt{3}/2$ is the Fermi velocity. Choosing
\begin{displaymath}
{\gamma_0} =
 \left(\begin{array}{c c}
{\sigma_z} & 0 \\
0 & {\sigma_z}
\end{array}\right),
\end{displaymath}
implies
\begin{equation}
{\gamma_1} =
 \left(\begin{array}{c c}
{\sigma_y} & 0 \\
0 & -{\sigma_y}
\end{array}\right),\quad
{\gamma_2} =
 \left(\begin{array}{c c}
{\sigma_x} & 0 \\
0 & {\sigma_x}
\end{array}\right).
\end{equation}
The two remaining anticommuting matrices may then be taken as
\begin{equation}
{\gamma_3} =
 \left(\begin{array}{c c}
0 & {\sigma_y} \\
{\sigma_y} & 0
\end{array}\right),\quad
{\gamma_5} = \left(\begin{array}{c c}
0 & -i{\sigma_y} \\
i{\sigma_y} & 0
\end{array}\right).
\end{equation}
This defines the "graphene representation" of $\gamma$-matrices.
\cite{herbut1}  ${\vec\sigma}$ are the standard Pauli matrices.

  Considering a more general model with further hoppings or weak anisotropies \cite{kohmoto} can be seen
not to destroy the Dirac points, but only to shift them in energy. As this can always be
compensated by a shift of the chemical potential,  the Lagrangian (2) provides the
low-energy description of the general free electronic Hamiltonian on a honeycomb lattice,
with the chemical potential tuned to the Dirac point.

  Note that the free Lagrangian besides the Lorentz symmetrt also possesses another
  chiral, pseudospin or "valley", global $SU_c (2)$ symmetry, generated by $\{ \gamma_3, \gamma_5, \gamma_{35} \}$.
  Both the Lorentz and the chiral symmetry of the free Lagrangian are
  emerging only at low energies, however, and the term quadratic in derivatives in $L_0$, for example, would spoil it. As will be shown shortly, both symmetries are also violated by the leading irrelevant quartic couplings introduced by the interactions.

Let us now consider the electron-electron interactions.
The Hamiltonian of a general four-fermion interaction has the form
\begin{equation}\label{Hint}
{H_{int}}={\sum_{\alpha,\beta,\gamma,\delta}}\langle\alpha\beta\mid{V}\mid\gamma\delta\rangle
{{r_\alpha}^\dagger}{{r_\beta}^\dagger}{r_\delta}{r_\gamma},
\end{equation}
where $r=u$ or $v$ are fermionic annihilation operators, and the matrix element corresponding to the
interaction potential $V(\vec{r})$ is given by
\begin{equation}
\langle\alpha\beta\mid{V}\mid\gamma\delta\rangle=\int{d\vec{x}}
{d\vec{y}} \varphi_\alpha ^*(\vec{x})
\varphi_\beta ^*(\vec{y}) {V(\vec{x}-\vec{y})}
{{\varphi_{\gamma}}(\vec{x})}{\varphi_{\delta}(\vec{y})}.
\end{equation}
Here we can take $\varphi_\alpha (\vec{x})$ to be a localized $p_z$-orbital
on the site $\alpha$, so that it belongs to either one of the two sublattices
of the honeycomb lattice. In general, there is no restriction on
the overlap of the wave-functions, and all the matrix elements
$\langle\alpha\beta\mid{V}\mid\gamma\delta\rangle$ are in principle
finite. Their relative sizes, however, may be rather different, and we discuss important simplifications
that follow in the limit of well localized orbitals in sec. VI.
In the following we will consider general "short-ranged"
interactions, which are defined by the interaction $V(\vec{x})$ with
a regular Fourier component $V(\vec{k}=0)$.
Without a loss in generality one may then take the interacting
Lagrangian for spinless fermions corresponding to $H_{int}$ as in Eq. (1)
where $M_1$ and $M_2$ are some constant $4\times4$ Hermitian matrices. The
interacting Lagrangian contains therefore at most $16 + (16 \times 15/2)=136$ independent real coupling
constants. However, the number of couplings in $L_{int}$ is severely reduced
 by the lattice symmetries, as we discuss next.

\subsection{Reflection symmetries}

Two obvious discreet symmetries of the honeycomb lattice that have not been broken by our choice of the Dirac points are the reflection symmetries through the lines A and B in Fig. 1. Let us consider the former symmetry first. It exchanges the two sublattices, but not the two Dirac points.  The
low-energy Lagrangian thus has to be invariant under the exchange of
the spinor components belonging to different sublattices, $u(\vec{k}){\leftrightarrow}v(\vec{k})$. Consequently, the  symmetry operator acting on the four-component Dirac spinor defined in Eq.\ (\ref{Dirac-spinor}) has the form
\begin{equation}
S=I\otimes\sigma_x=\gamma_2.
\end{equation}
Since under this reflection $q_x \rightarrow q_x$ and $q_y \rightarrow -q_y$, $L_0$ is evidently invariant under $S$.
The invariance of $L_{int}$ under this reflection symmetry requires both
matrices $M_1$ and $M_2$ to either commute or anticommute with the operator $S$:
\begin{equation}
[S,M_1]=[S,M_2]=0,
\end{equation}
or
\begin{equation}
\{S,M_1\}=\{S,M_2\}=0.
\end{equation}

Similarly, the reflection symmetry through the line $B$ exchanges the two Dirac points, while not exchanging the sublattice labels. It corresponds therefore to
 \begin{equation}
T=i{\gamma_1}{\gamma_5}=\left(\begin{array}{c c}
0 & I_2 \\ I_2 & 0
\end{array}\right).
\end{equation}
Recalling that under this transformation $q_x \rightarrow -q_x$ and $q_y \rightarrow q_y$, it is evident that
the free Lagrangian in graphene representation is indeed invariant under $T$ as well. Demanding the interacting Lagrangian $L_{int}$ to be invariant under the action of the operator $T$ on the Dirac spinor, implies that both matrices $M_1$ and $M_2$ either commute or anticommute with $T$ as well. In other words, both
matrices $M_1$ and $M_2$ have to be either even or odd with respect to $T$:
\begin{equation}
[T,M_1]=[T,M_2]=0,
\end{equation}
  or
\begin{equation}
\{T,M_1\}=\{T,M_2\}=0.
\end{equation}
Together with the combination of the two reflections S and T, and the identity operation, the two symmetry operations form the {\it dihedral group} (or Klein's {\it vierergrouppe}, in older literature) $D_2$: $D_2= \{ 1, S, T, ST \} = Z_2 \times Z_2$, the symmetry group of a rectangle. Note that the transformation $ST$ is just the space inversion, and that the rotation by $\pi/2$ {\it does not} belong to $D_2$.

One may now classify all the four-dimensional matrices into four categories, according to their transformation
under the two reflection  symmetries $S$ and $T$, respectively: even-even, $A\equiv \{I,\gamma_2,i\gamma_0\gamma_3,i\gamma_1\gamma_5 \}$, even-odd,
$B\equiv\{i\gamma_0\gamma_1, \gamma_{35}, i\gamma_0\gamma_5,i\gamma_1\gamma_3 \}$,
odd-even, $C\equiv\{\gamma_0,i\gamma_0\gamma_2,\gamma_3,i\gamma_2\gamma_3\}$,
and odd-odd, $D\equiv\{\gamma_1,i\gamma_1\gamma_2,\gamma_5,i\gamma_2\gamma_5\}$.
The interacting Lagrangian symmetric under the $D_2$
is thus restricted to be of the following form
\begin{widetext}
\begin{equation}\label{Lagrangian}
L_{int}={a_{ij} }({\Psi^\dagger}A_i\Psi)({\Psi^\dagger}A_j\Psi)+
{b_{ij}}({\Psi^\dagger}{B_i}\Psi)({\Psi^\dagger}{B_j}\Psi)
+{c_{ij}}({\Psi^\dagger}{C_i}\Psi)({\Psi^\dagger}{C_j}\Psi)
+{d_{ij}}({\Psi^\dagger}{D_i}\Psi)({\Psi^\dagger}{D_j}\Psi),
\end{equation}
\end{widetext}
where $o_{ij}$, $o=a,b,c,d$, $i,j=1,...,4$, are real and symmetric. The maximal number of independent real parameters specifying the allowed couplings is thus already reduced to forty, since each $o_{ij}$ has ten independent components.

\subsection{Translational invariance}

  The generator $\gamma_{35}= \sigma_z \otimes I_2 $ of the chiral symmetry plays a special role. It is in fact
the generator of translations. To see this recall that under a translation by $\vec{R}$  the electron fields transform as
\begin{equation}
r(\vec{k}, \omega) \rightarrow e^{i \vec{k}\cdot \vec{R} } r(\vec{k},\omega)
\end{equation}
where $r=u,v$. The Dirac field under the same transformation thus changes as
\begin{equation}
\Psi(\vec{q}, \omega) \rightarrow e^{ i (\vec{K} \cdot \vec{R}) \gamma_{35} }  e^{i\vec{q}\cdot \vec{R}} \Psi(\vec{q}, \omega),
\end{equation}
or, in real space,
\begin{equation}
\Psi(\vec{x},\tau) \rightarrow e^{ i (\vec{K} \cdot \vec{R}) \gamma_{35} } \Psi(\vec{x}+\vec{R},\tau).
\end{equation}
Translational invariance requires therefore that $L_{int}$ is a scalar under the transformations generated by
$\gamma_{35}$, which we will denote as $U_c (1)$. It is easy to see that this is precisely the same as demanding the conservation of momentum in the interaction terms. The reader is also invited to convince herself that the terms with the higher-order derivatives in $L_0$ would also be invariant under the $U_c (1)$.

  First, we observe that there are eight linearly independent bilinears that are {\it scalars} under the $U_c (1)$:
  \begin{equation}
  X_{Fi}= \Psi^\dagger F_i \Psi,
  \end{equation}
  where $F=A,B,C,D$ and $i=1,2$. The remaining eight bilinears can be grouped into four {\it vectors} under the same $U_c (1)$:
  \begin{equation}
  \vec{\alpha} = (\Psi^\dagger A_3 \Psi, \Psi ^\dagger  B_3 \Psi),
  \end{equation}
   \begin{equation}
  \vec{\beta} = (\Psi^\dagger B_4 \Psi, \Psi ^\dagger A_4 \Psi),
  \end{equation}
\begin{equation}
  \vec{\gamma} = (\Psi^\dagger C_3 \Psi, \Psi ^\dagger D_3 \Psi),
  \end{equation}
\begin{equation}
  \vec{\delta} = (\Psi^\dagger C_4 \Psi, \Psi ^\dagger D_4 \Psi),
  \end{equation}

  The invariance under $U_c (1)$ implies therefore that the interacting Lagrangian has the following form
\begin{eqnarray}
L_{int} = \sum_{Fi}  g_{Fi} X_{Fi} ^2  + \sum_{F} g_{F} X_{F1} X_{F2}  \\ \nonumber
 + g_{\alpha \beta}  \vec{\alpha}\times \vec{\beta} + g_{\gamma \delta}  \vec{ \gamma} \cdot \vec{\delta}
 + \sum_{\rho=\alpha,\beta,\gamma,\delta} g_\rho \vec{\rho}\cdot \vec{\rho}.
\end{eqnarray}
The number of possible independent couplings is down to eighteen.

\subsection{Time-reversal}

The set of the allowed couplings is further reduced by the time-reversal symmetry. The microscopic interacting Hamiltonian (\ref{Hint}) is invariant
under the time-reversal, and therefore the corresponding
low-energy Lagrangian has to possess the same invariance.
The time-reversal symmetry requires that
${I_t}H{I_t}^{-1}=H$, where $I_t$ is the antiunitary operator
representing the time-reversal symmetry, and thus has the form
${I_t}=U K$, with $U$ representing the unitary part of $I_t$ and
$K$ is the complex conjugation. To find the form of $I_t$
 let us consider first the {\it massive} Dirac Hamiltonian
\begin{equation}\label{H0m}
H=i{\gamma_0}{\gamma_i}{p_i}+ {m_1}{\gamma_0},
%+ {m_1}{i\gamma_0}{\gamma_3}+
%{m_2}{i\gamma_0}{\gamma_5},
\end{equation}
with the mass $m_1$ describing the imbalance in the chemical
potential on the two sublattices. \cite{semenoff} Recalling that momentum changes sign
under the time-reversal, ${I_t}{p_i}{I_t}^{-1}=-{p_i}$, in the graphene representation
the invariance of the above Hamiltonian under the same transformation implies
\begin{eqnarray}
\{U,i{\gamma_0}{\gamma_1}\}=[U,i{\gamma_0}{\gamma_2}]=[U,\gamma_0]=0,
\end{eqnarray}
and hence $U= i e^{i\phi} \gamma_1 (\cos\theta \gamma_3 + \sin\theta \gamma_5)$.
Within the simplest framework of the tight-binding model with uniform hopping the time-reversal
operator is not uniquely determined. We thus consider
a generalized tight-binding model with anisotropic hopping defined as
\begin{equation}
\label{anisotropic}
H_{aniso}={\sum_{\vec{A}, i}}(\tilde{t}+\delta{\tilde{t}_{\vec{A},i}}) u^\dagger (\vec{A)}  v (\vec{A}+\vec{b_i})
+ H.c.,
\end{equation}
where
\begin{equation}
\delta{\tilde{t}_{\vec{A},i}}=
\frac{1}{3}{\Delta}(\vec{A})e^{i\vec{K}\cdot{\vec{b_i}}}e^{i\vec{G}\cdot{\vec{A}}}+
c.c.
\end{equation}
represents a non-uniform hopping, and
$\vec{G}=2 \vec{K}$. \cite{hou}  On a lattice, this particular set of hoppings generates the
so-called Kekule texture. Near the two Dirac points the Hamiltonian $H_{aniso}$ reads
\begin{equation} \label{kekule}
H_{aniso}=i{\gamma_0}{\gamma_i}{p_i}+{m_2}i{\gamma_0}{\gamma_5}+{m_3}i{\gamma_0}{\gamma_3},
\end{equation}
where ${m_2}=Im(\Delta(\vec{r}))$ and ${m_3}=Re(\Delta(\vec{r}))$.
The two masses $m_2$ and $m_3$ therefore provide the low-energy representation of a
completely real microscopic Hamiltonian, so that we postulate that
$H_{aniso}$ is also time-reversal symmetric. In graphene representation this requires
the unitary part of the time-reversal operator to obey the following algebra:
\begin{equation}\label{condT}
[U,i{\gamma_0}{\gamma_3}]=\{U,i{\gamma_0}{\gamma_5}\}=0.
\end{equation}

The matrix $T=i{\gamma_1}{\gamma_5}$ satisfies conditions
(\ref{condT}) and thus the unitary part of the operator $I_t$
acting on the spinless Dirac field (\ref{Dirac-spinor}) is \cite{herbutT}
\begin{equation}
U=T= i{\gamma_1}{\gamma_5}=\left(\begin{array}{c c}
0 & I_2 \\
I_2 & 0
\end{array}\right),
\end{equation}
with $I_2$ as the $2\times2$ unity matrix. The unitary part of the
time-reversal operator thus simply exchanges the components of the Dirac
spinor $\Psi({\vec x},\tau)$ with different valley indices, as expected.
It also happens to be the same as one of the two matrices representing the
reflection operators from $D_2$.

  Another way of arriving at the same form for the time-reversal operator is
to postulate that an arbitrary chiral transformation of the Dirac Hamiltonian in Eq. (29)
yields a time-reversal invariant Hamiltonian. Alternatively, our derivation may be understood as
a demonstration of commutativity of the chiral and time-reversal transformations.

Since we have already used the invariance under $T$ to restrict the interacting Lagrangian, time-reversal
invariance will be observed if the remaining terms are even under complex conjugation. All
the terms $X_{Fi} ^2$ and $\vec{\rho} \cdot \vec{\rho}$ are thus automatically invariant under time-reversal,
but among the remaining six mixed terms, the terms  $X_{C1} X_{C2}$, $X_{D1} X_{D2}$, and $\vec{\gamma}\cdot \vec{\delta}$ are odd. Time-reversal invariance implies therefore that
\begin{equation}
g_C = g_D =g_{\gamma \delta}=0,
\end{equation}
which leaves then at most fifteen independent couplings.

\section{Enlargement of symmetry}

We found that the exact symmetries of the microscopic Hamiltonian, $D_2$, translational, and the time-reversal, leave  at most fifteen independent short-range couplings. Anticipating some of the results, it is interesting to deduce the further reductions of the number of couplings if one by hand imposes larger symmetries onto the interaction Lagrangian $L_{int}$.

\subsection{Rotational invariance}

  Since the rotation by $\pi/2$ is not a member of the $D_2$, the matrices $\gamma_1$ and $\gamma_2$ appear asymmetrically in $L_{int}$. If we demand that they appear symmetrically, $L_{int}$ becomes fully rotationally invariant. This is achieved if
  \begin{equation}
  g_A = g_B= g_{\alpha \beta} = g_{A2}-g_{D1}= g_{B1}-g_{C2}= g_\beta - g_\delta =0.
  \end{equation}
  Let us call the interacting Lagrangian with the rotational invariance imposed this way $L_{int, rot}$. It would be described by at most nine couplings.

  \subsection{Lorentz invariance}

 Imposing further the Lorentz invariance would require that on top of the above restrictions one also has that
 \begin{equation}
 g_{A1} + g_{B1}= g_{A2} +g_{B2} = g_\gamma +  g_\beta=0.
 \end{equation}
 With the restrictions in the previous two equations the Lagrangian has only six couplings constants, and may be cast in a manifestly Lorentz invariant form, worth displaying:
 \begin{eqnarray}
 L_{int, lor} = g_{A1} (\bar{\Psi} \gamma_\mu \Psi)^2  + g_{B2} (\bar{\Psi} \gamma_\mu \gamma_{35} \Psi)^2 + \\ \nonumber
 g_{C1} (\bar{\Psi} \Psi)^2 + g_{D2} ( \bar{\Psi} \gamma_{35} \Psi)^2
 + g_\alpha  [(i \bar{\Psi} \gamma_3 \Psi)^2 +  \\ \nonumber
 (i \bar{\Psi} \gamma_5 \Psi)^2 ] +
 g_\gamma [ (\bar{\Psi} \gamma_\mu \gamma_3 \Psi)^2 + (\bar{\Psi} \gamma_\mu \gamma_5 \Psi)^2 ].
 \end{eqnarray}

\subsection{Chiral invariance}

Finally, the maximally invariant interacting Lagrangian would be with the full, i. e.  both the Lorentz and the chiral, symmetry of the non-interacting part. This is achieved by setting in the last equation
\begin{equation}
g_{C1}-g_\alpha=g_{B2}-g_\gamma=0.
\end{equation}
The interacting Lagrangian can in this case be written as
\begin{equation}
L_{int,max} = g_{A1} S_\mu ^2 + g_{D2} S^2 + g_{C1} \vec{V}^2 + g_{B2} \vec{V}_\mu ^2,
\end{equation}
where the participating bilinears in Dirac fields,
\begin{equation}
S_\mu= \bar{\Psi} \gamma_\mu \Psi,
\end{equation}
\begin{equation}
S= \bar{\Psi} \gamma_{35} \Psi,
\end{equation}
\begin{equation}
\vec{V}=( \bar{\Psi} \Psi, i \bar{\Psi} \gamma_3 \Psi,i \bar{\Psi} \gamma_5 \Psi ),
\end{equation}
\begin{equation}
\vec{V}_\mu = ( \bar{\Psi} \gamma_\mu \gamma_{35} \Psi, \bar{\Psi} \gamma_\mu \gamma_{3} \Psi, \bar{\Psi} \gamma_\mu \gamma_{5} \Psi),
\end{equation}
are the scalar (vector), scalar (scalar), vector (scalar), and vector (vector) under the chiral (Lorentz) transformation. The last form makes the Lorentz and the chiral symmetry of the Lagrangian $L_0 + L_{int,max}$ manifest. Such a {\it maximally symmetric} Lagrangian contains therefore at most only four coupling constants.

\section{Fierz transformations}

The number of independent couplings is further reduced by the existence of algebraic identities between seemingly different quartic terms. The derivation of the so-called Fierz transformation, which allows one to write a given local quartic term in terms of other quartic terms is provided in Appendix A.  A systematic application of this transformation allows one to reduce the number of independent couplings for a given symmetry of the interacting Lagrangian.

\subsection{General problem}

   The application of Fierz identity to the set of quartic terms allowed by the assumed symmetry in principle leads to
the set of linear constraints of the form
\begin{equation}
F X =0,
\end{equation}
where $F$ is a real typically asymmetric matrix, and $X$ is a column; the elements of which are the quartic terms allowed by the symmetry. Of course, only the quartic terms which share the same symmetry may be related by Fierz transformations. For example, in the maximally symmetric case $X^\top = (S^2, S_\mu ^2, \vec{V}^2, \vec{V}_\mu ^2)$. When the number of couplings is small it is easy to discern the linearly independent combinations of the original terms, but when it is not, as the case is for $D_2 \times U_c (1) \times I_t$ microscopic symmetry of $L_{int}$, one needs a more general method of doing so.

In Appendix B we show that an asymmetric matrix such as the Fierz matrix $F$ can be written in the diadic form \cite{musicki} as
\begin{equation}
 F= \sum_{i} \mu_i^{1/2} | \nu_i \rangle \langle \mu_i |
 \end{equation}
 where $\{ \mu_i \}$ is the real spectrum of the related symmetric matrix $S_F =F^\top F$. In the eigenbasis of $S_F$ we can write, in Dirac notation,
 \begin{equation}
 | X \rangle = \sum_i |\mu_i \rangle\langle \mu_i | X\rangle,
 \end{equation}
 so that the above linear equations can be written as
 \begin{equation}
 F X = \sum_i \mu_i^ {1/2}   \langle \mu_i | X \rangle  |\nu_i\rangle =0.
 \end{equation}
 Since the vectors $\{ |\nu_i \rangle \}$ also form a basis, it must be that either:

  \noindent
  a) for $\mu_i \neq 0$,  $\langle \mu_i | X \rangle=0$ , or

  \noindent
  b) $\mu_i=0$, so that $ \langle \mu_i | X \rangle \neq 0$.

The first set provides us then with the linearly independent constraints, and the second  with the set of remaining
 linearly independent quartic terms. Since the matrices $F$ and $S_F$ obviously have the same kernels, the number of independent coupling constants allowed by the symmetry is simply the dimension of the kernel of the appropriate Fierz matrix.

 \subsection{Maximally symmetric case}

 Let us consider the simplest example of the quartic term with the full Lorentz and chiral symmetry, $L_{int,max}$ first. Defining the vector $X$ as in the above leads to the Fierz matrix
\begin{equation}
F =
 \left(\begin{array}{c c c c }
3 & 3 & 3 & -1 \\
5 & 1 & 1 & 1 \\
3 & 3 & 3 & -1 \\
9 & -3 & -3 & 5
\end{array}\right),
\end{equation}
with the two-dimensional kernel with the zero-eigenstates
\begin{equation}
\langle \mu_1 |= \frac{1}{\sqrt{2}} (0,-1,1,0),
\end{equation}
\begin{equation}
\langle \mu_2 |= \frac{1}{2\sqrt{3}} (-1,1,1,3),
\end{equation}
and $\mu_1 = \mu_2 =0$. The remaining two eigenvalues are $\mu_3=64$, and $\mu_4 = 144$. The general method explained above implies that the general maximally symmetric interacting Lagrangian can be written as
\begin{equation}
L_{int,max} = \lambda_1 (\vec{V}^2 - S_\mu ^2) + \lambda_2 (-S^2 + S_\mu ^2 +\vec{V}^2 + 3 \vec{V}_\mu ^2),
\end{equation}
with the "physical" couplings
\begin{equation}
\lambda_1 = \frac{ g_{C1} -g_{A1} }{2},
\end{equation}
\begin{equation}
\lambda_2 = \frac{- g_{D2}+ g_{A1}+ g_{C1}  + 3 g_{B2}  }{12}.
\end{equation}
The two remaining linearly independent combinations vanish due to Fierz identity. So the maximally symmetric interacting theory is specified by only two quartic coupling constants, which may be chosen to be any linearly independent combinations of the above $\lambda_1$ and $\lambda_2$.

\subsection{Lorentz-symmetric case}

  We may then proceed to find the independent couplings for the next case in order in complexity, $L_{int, lor}$, with the chiral symmetry broken down to $U_c (1)$. If we define the six-dimensional vector
  \begin{equation}
  X^\top= ( S^2, S_\mu ^2, V_1 ^2, V_2 ^2 + V_3 ^2 , S_{1 \mu} ^2, S_{2 \mu}^2 + S_{3 \mu} ^2 )
  \end{equation}
  the Fierz matrix is found to be
  \begin{equation}
F =
 \left(\begin{array}{c c c c c c}
1 & 1 & 5 & -1 & 1 & -1\\
3 & 3 & -1 & 5 & -7/3 & -1/3\\
3 & 3 & 3 & 3 & -1 & -1 \\
5 & 1 & 1 & 1 & 1 &1 \\
3 & -1 & 3 & -3 & 3 & 1\\
3 & -1 & -3 & 0 & 1 & 2
\end{array}\right),
\end{equation}
 with the three-dimensional kernel spanned by
 \begin{equation}
 \langle \mu_1 | = \frac{1}{5 \sqrt{2}} (-2,3,1,0,0,6),
 \end{equation}
 \begin{equation}
 \langle \mu_2 | = \frac{1}{5 \sqrt{2}} (-1,4,-2,0,5,2),
 \end{equation}
 \begin{equation}
 \langle \mu_3 | = \frac{1}{2 \sqrt{55}} (-3,-7,3,10,7,2).
 \end{equation}
 $L_{int,lor}$ can now be written in terms of only three linearly independent quartic terms, in complete analogy with the maximally symmetric case.

 \subsection{Lower symmetries}

  The symmetry ladder may now be climbed back to the $D_2 \times U_c (1) \times I_t$ minimally symmetric Lagrangian. First, reducing Lorentz to rotational symmetry increases the number of independent couplings to four. Removing the rotational symmetry finally increases the number of couplings to six. We may note in passing that there are two independent Fierz identities between the mixed terms that obey the time-reversal symmetry in Eq. (28):
  \begin{equation}
  3 X_{A1} X_{A2} - X_{B1} X_{B2} + \vec{\alpha} \times \vec{\beta} =0,
  \end{equation}
  \begin{equation}
  X_{A1} X_{A2}+ X_{B1} X_{B2} + \vec{\alpha} \times \vec{\beta}=0,
  \end{equation}
  so that the three mixed terms in fact contribute a single independent coupling.

  \section{Renormalization group}

    Having determined the independent coupling constants for each symmetry, we now proceed to study their changes with the decrease of the upper cutoff $\Lambda$. We will be particularly interested in fixed points of such renormalization group transformations, as they will provide the information on the quantum metal-insulator transitions that can be induced by increase in interactions.

   \subsection{Maximally symmetric theory}

  Let us again begin with the maximally symmetric Lagrangian, $L= L_0 + L_{int,max}$. There are only two coupling constants to consider in this case, and we choose them to be $g_{D2}$ and $g_{A1}$, which correspond to $S^2$ and $S_\mu ^2$ quartic terms, respectively. If any of the other two terms would become generated by the renormalization  transformation we would use the Fierz identity to rewrite it in terms of $S^2$ and $S_\mu ^2$. Alternatively, one may wish to renormalize the theory as written in terms of physical couplings in Eq. (53). This procedure, completely equivalent to what is pursued here, is described in Appendix C.  As we integrate the fermionic modes lying in the 2+1 dimensional momentum shell \cite{remarkshell} from $\Lambda/b $ to $\Lambda$, with $b>1$, to quadratic order in  coupling constants we find
 \begin{equation}
 \frac{d g_{D2}}{d \ln b} = -g_{D2}  - g_{D2}^2   + 2 g_{A1}^2  + 3 g_{D2} g_{A1},
 \end{equation}
 \begin{equation}
 \frac{d g_{A1}}{d \ln b} = -g_{A1} + g_{A1}^2 + g_{D2} g_{A1}.
 \end{equation}
 We rescaled the couplings here as $2 g \Lambda/\pi^2 \rightarrow g$.
 To this order no other types of quartic terms actually get generated, and the Fierz transformation turns out not to be  necessary. The limit of the above equations that survives the extension to a large number of Dirac fields also agrees with the previous calculation. \cite{kaveh}

    The above flows, besides the Gaussian, exhibit three fixed points at finite couplings (Fig. 3). The first critical point (A) is at $ g_{D2}= -1$, $g_{A1}=0$, and the second critical point (C) is at $g_{D2}=(\sqrt{5}-1)/2 $, $g_{A1}= (3-\sqrt{5})/2$. There is also a bicritical fixed point (B) that separates the domains of attraction of the two critical points, at $g_{D2}=-(\sqrt{5}+1)/2 $, $g_{A1}=(\sqrt{5}+3)/2 $.

 \begin{figure}[t]
{\centering\resizebox*{85mm}{!}{\includegraphics{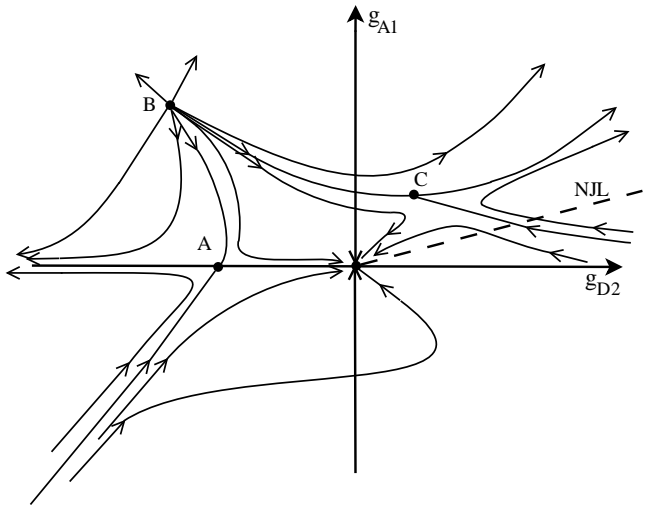}}
\par} \caption[] {Schematic flow diagram of the two coupling constants in the maximally symmetric theory. The fixed point A describes the continuous transition into a time-reversal symmetry broken insulator, and C the dynamical generation of the chiral symmetry breaking mass. The line $g_{D2}=0$ describes the Thirring model, and the dashed line the Nambu-Jona-Lasinio  model in 2+1 dimensions. (See the text.)}
\end{figure}

 The physical interpretation of the critical point A is obvious. Since we can tune through it by keeping $g_{A1}=0$ and increasing $g_{D2}$ over a certain negative value, it should describe the transition into the insulator with the gap that breaks the time-reversal symmetry, described by
 \begin{equation}
 \langle \bar{\Psi} \gamma_{35} \Psi \rangle \neq 0.
 \end{equation}
 This state is obviously favored at a large and negative $g_{D2}$. Note that since $\gamma_{35}$ commutes with $\gamma_\mu$, the line $g_{A1}=0$ is invariant under RG. In fact, the perturbative $\beta$-function along this line has to be identical as the one in the Gross-Neveu model. We can therefore simply use the already existing higher-order estimates \cite{vasilev, gracey} and the numerical results \cite{karkkainen, wetterich} to find the critical exponents describing this particular metal-insulator transition. We return to this fixed point shortly.

   The physical interpretation of the critical point C is less obvious, but we can think of it as follows. First, note that the line $g_{D2}=0$, $g_{A1}>0$, which describes the Thirring model, \cite{christofi} belongs to the domain of attraction of C. Also, the Fierz transformations in Eq. (50) imply
   \begin{equation}
   g_{D2} S^2 + g_{A1} S_\mu ^2 = (g_{D2}-2 g_{A1}) S^2 - g_{A1} \vec{V}^2 .
   \end{equation}
   The line $g_{D2}-2 g_{A1}=0$ for $g_{A1}>0$ , which we name the Nambu-Jona-Lasinio (NJL) line, \cite{nambu} also falls into the domain of attraction of the critical point C. Along this line, however, there should be a transition into an insulating state with
   \begin{equation}
   \langle \vec{n} \cdot \vec{V} \rangle \neq 0,
   \end{equation}
   where $\vec{n}$ is a unit vector. Such a state is clearly favored at a large and positive $g_{A1}$ along the NJL line, and breaks the chiral $SU_c (2)$ symmetry down to $U(1)$. We therefore identify C as the metal-insulator critical point governing the chiral-symmetry breaking transition in {\it both} Thirring and NJL models with a single Dirac field.

     The picture suggested by the above one-loop calculation in the maximally symmetric theory appears quite natural.  There are two possible insulating phases, each breaking either chiral or time-reversal symmetry, which correspond to possible "masses" for the Dirac fermions. Both metal-insulator transitions are continuous, and are described by different critical points.

Of course, the true low-energy theory on the honeycomb lattice is much less symmetric than the one studied in this section. Nevertheless, we will argue that the two identified critical points may in fact be stable at least with respect to weak manifest breaking of the Lorentz and chiral symmetries.

 \subsection{Broken Lorentz symmetry}

   The explicit breaking of Lorentz symmetry down to the rotational symmetry can be easily implemented by adding to $L_{int, max}$ a small symmetry breaking term
   \begin{equation}
   \delta (\bar{\Psi} \gamma_0 \Psi)^2,
   \end{equation}
   which by virtue of being only rotational symmetric is guaranteed to be linearly independent of $S^2$ and $S_\mu ^2$. A weak perturbation $\delta$ to the lowest order in couplings $g_{D2}$ and $g_{A1}$ then flows according to
   \begin{equation}
   \frac{d \delta}{d\ln b} = \delta ( -1 - g_{D2}+ g_{A1}) + O(\delta^2).
   \end{equation}
   We thus find the critical point C to be stable with respect to weak Lorentz symmetry breaking to one loop, the bicritical point B to be unstable, and A marginal. We suspect that this result, although clearly an outcome here of an uncontrolled approximation, may be indicative of the true state of affairs. Hereafter we will assume that the critical points A and C are stable with respect to weak breaking of the Lorentz symmetry in the Lagrangian. It may also be worth mentioning that the complete one-loop $\beta$-functions for $\delta$, $g_{A1}$ and $g_{D2}$, which we have computed but have not shown, do not lead to any new critical points at $\delta\neq 0$.

 \subsection{Broken chiral symmetry}

   The simplest quartic term with the full Lorentz symmetry, and only $U_c (1)$ subgroup of the full chiral symmetry
   may be written as
   \begin{equation}
   L_{int, lor}= g_{D2} S^2 +  g_{C1} V_1 ^2 + g_{\alpha} (V_2 ^2 + V_3 ^2).
   \end{equation}
   The Fierz transformation matrix given above implies that these three quartic terms are indeed linearly independent. When $g_{C1}= g_\alpha$, the Lagrangian  $L_{int,lor}$ acquires the full chiral $SU_c(2)$ symmetry, and may be rewritten as $L_{int,max}$.

 Using the Fierz transformation, and after a convenient rescaling of the couplings as $g\Lambda /3 \pi^2 \rightarrow g$, to the quadratic order one finds
 \begin{eqnarray}
 \frac{d g_{D2}}{d \ln b} = -g_{D2}  - 6 g_{D2}^2 - \\ \nonumber
 4 g_\alpha ^2
 + 6 g_{D2} g_{C1} +12 g_{D2} g_\alpha -8 g_{C1} g_\alpha,
 \end{eqnarray}
 \begin{equation}
 \frac{d g_{C1}}{d \ln b} = -g_{C1} -6 g_{C1}^2 -8 g_\alpha ^2 + 6 g_{D2} g_{C1} - 4 g_{C1} g_\alpha,
 \end{equation}
 \begin{equation}
 \frac{d g_{\alpha}}{d \ln b} = -g_{\alpha} -8 g_{\alpha}^2 +6 g_{D2} g_\alpha  -10 g_\alpha g_{C1}.
 \end{equation}

 The two chirally symmetric critical points from the previous section now appear at $g_{D2}=-1/6$, $g_{C1}=g_\alpha=0$ (A), and $g_{D2}= (3\sqrt{5}-7)/12$, $g_{C1}=g_\alpha=(\sqrt{5}-3)/12$ (C), and both remain critical, even in absence of chiral symmetry in the Lagrangian. There is, however, an additional critical point (E) at $g_{D2}= (\sqrt{5}-2)/6$, $g_{C1}= -2 g_\alpha = (\sqrt{5}-3)/6$. Note also that the plane $g_\alpha =0$ is invariant under the renormalization group, but whereas the fixed point A in that plane is critical, the fixed point (D) at $g_{D2}=0$, $g_{C1}=-1/6$ is bicritical. One also finds that the line $g_{D2}=0$, $g_{C1}<0$, $g_\alpha$-infinitesimal and positive intersects the critical surface which contains the point E, whereas for $g_\alpha$-infinitesimal and negative the critical behavior is governed by C. For weak $g_\alpha$ therefore there is a crossover from the fixed point at D toward either C or E, depending on the sign. Interestingly, for negative $g_\alpha$ chiral symmetry becomes fully restored at the transition, at least within our one-loop calculation. \cite{calabrese}

\section{Atomic limit}

  Motivated by the one-loop results, we will assume hereafter that the Lorentz symmetry becomes restored at long distances in the domain of interest, and that we need only consider $L_{int, lor}$ with the three couplings from the last section. The situation however, can then be simplified even further, as we discuss in this section.

   Consider the interaction Hamiltonian in Eq. (11). If the $p_z$-orbitals are well localized on their corresponding lattice sites, we may neglect the matrix elements with $\alpha\neq \gamma$ or $\beta\neq \delta$. Keeping only the remaining, dominant matrix elements then one obtains the "atomic limit" of the general interaction Hamiltonian
   \begin{equation}
   H_{int}\rightarrow H_{lat} = \sum_{\alpha,\beta} V_{\alpha, \beta} n_{\alpha} n_\beta
   \end{equation}
   where $n_\alpha$ is the electron number operator at site $\alpha$. The class of Hamiltonians $H_{lat}$ is evidently still rather broad, and would for example include all lattice interacting Hamiltonians.

   Writing the lattice Hamiltonian $H_{lat}$ in terms of the Dirac fields, however, imposes yet another restriction on the coupling constants. Since $H_{lat}$ is written in terms of lattice-site particle number operators, any Dirac quartic term evidently must contain an equal number of $u^\dagger$ ($v^\dagger$) and $u$ ($v$) fields. On the other hand, the $g_\alpha $-term from above in momentum space can be written schematically as
   \begin{equation}
  (V_2 ^2 + V_3 ^2) \sim  ( u_1 ^\dagger v_2 + v_1 ^\dagger u_2)(u_2 ^\dagger v_1 + v_2 ^\dagger u_1)
  \end{equation}
  and thus contains the terms forbidden in the atomic limit,\cite{remarkatomic} such as $u_1 ^\dagger v_2 u_2 ^\dagger v_1$. The index 1 and 2 refers here to the two Dirac points. This implies that for any lattice Hamiltonian $H_{lat}$  we must have
  \begin{equation}
  g_\alpha =0.
  \end{equation}
  Note that the plane $g_\alpha=0$ is invariant under the change of cutoff in the above one-loop calculation. It is easy to see that this feature of the $\beta$-functions for $g_{D2}$, $g_{C1}$ and $g_\alpha$ is in fact true to all orders in perturbation theory. The matrices $\gamma_{35}$ and $I$ in the remaining two terms in $L_{int,lor}$ commute with the Dirac propagator, and therefore an arbitrary diagram containing $g_{D2}$ and $g_{C1}$ terms can contribute only to the renormalized $g_{D2}$ and $g_{C1}$ couplings. So imposing $g_\alpha=0$ at an arbitrary cutoff guarantees its vanishing at all others.

     It is therefore not only physically justified but also internally consistent to consider only the two couplings $g_{D2}$ and $g_{C1}$ in the Lorentz symmetric, but chirally asymmetric low-energy theory. The one-loop result in this plane is depicted in Fig. 4. The transition is either into the time-reversal-symmetry-broken, or into chiral-symmetry-broken insulator. Several features of this flow diagram that should be generally valid are worth mentioning.

     1) There should be two critical points, both unstable in a single direction. Bicriticality of A, for example, would imply that the transition for negative $g_{D2}$ at a weak positive $g_{C1}$ is first order. This, however, seems unlikely on physical grounds, and, also, it is not found in the explicit large-N generalization of the theory, when the two $\beta$-functions are known to decouple. \cite{kaveh}

     2) The two critical points have identical critical behavior. This is because the term $\bar{\Psi} \gamma_{35} \Psi$, in graphene representation, under the transformation
     \begin{equation}
     \Psi \rightarrow \frac{1}{2} [ i (I_2 + \sigma_z)\otimes \sigma_z  + (I_2 - \sigma_z)\otimes I_2]  \Psi,
     \end{equation}
     \begin{equation}
     \Psi^\dagger \rightarrow \frac{1}{2}  \Psi ^\dagger [ i (I_2 + \sigma_z)\otimes \sigma_z  + (I_2 - \sigma_z)\otimes I_2],
     \end{equation}
     goes into $\bar{\Psi} \Psi$, and vice versa, while $L_0$ remains invariant. This also means that the two $\beta$-functions are symmetric under the exchange $g_{D2}\leftrightarrow  g_{C1}$. Both critical points are thus in the universality class of the Gross-Neveu model.

     3) At the line $g_{D2}= g_{C1}$ the single $\beta$-function becomes
     \begin{equation}
     \frac{d g_{D2}}{d \ln b}=- g_{D2},
     \end{equation}
     i. e. $g_{D2}$ flows according solely to its canonical dimension. This is because
     \begin{equation}
     (\bar{\Psi} \gamma_{35} \Psi)^2 + ( \bar{\Psi} \Psi)^2  = 2(\Psi^\dagger _+ \sigma_z \Psi_+)^2 + 2 (\Psi^\dagger _- \sigma_z \Psi_-)^2,
     \end{equation}
     where $\Psi^\dagger = (\Psi_+ ^\dagger, \Psi_- ^\dagger)$. Since all $\gamma_\mu$ are block-diagonal, $+\vec{K}$ and $-\vec{K}$ components at this line decouple. The partition function factorizes into a product of two Gross-Neveu partition functions, each containing a single {\it two-component} Dirac fermion. Along this line the system is believed to have the  metal-insulator transition, possibly continuous, \cite{wetterich} but the $\beta$-function vanishes at least to the order $g_{D2}^3$. \cite{gracey}

        \begin{figure}[t]
{\centering\resizebox*{85mm}{!}{\includegraphics{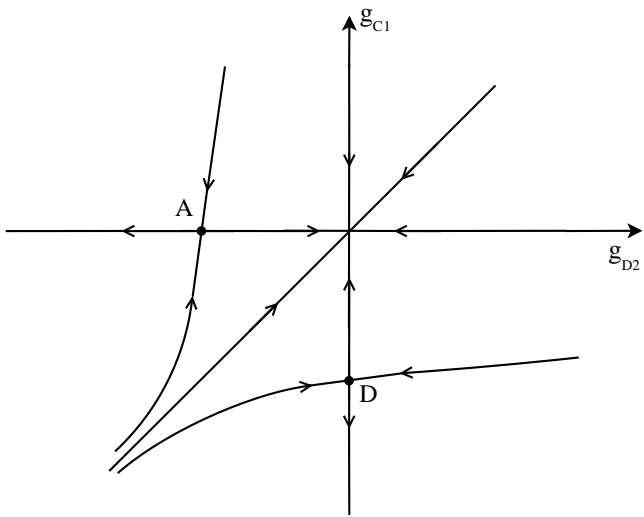}}
\par} \caption[] {The flow diagram in the $g_{D2}-g_{C1}$ plane. The possible non-perturbative fixed point at the $g_{D2}=g_{C1}$ line, which governs the transition in the two-component Gross-Neveu theory is not shown.}
\end{figure}

\section{Critical exponents}

 Each identified metal-insulator transition is characterized by a set of critical exponents. We will here focus on the three already mentioned in the introduction: the correlation length exponent $\nu$, the dynamical exponent $z$, and the Dirac fermion anomalous dimension $\eta_\Psi$. The other exponents can then be obtained from the usual scaling relations. \cite{book}

   First, since all the identified critical points exhibit Lorentz symmetry,
   \begin{equation}
   z=1.
   \end{equation}
 We also find that the exponent $\nu$ is unity at all critical points as well, but this is clearly an artifact of the one-loop calculation. In general, $\nu$ is expected to be different at different critical points. The same goes for $\eta_\Psi$, which vanishes in one-loop calculation, but will be finite in general.

  In the atomic limit, when under the assumed Lorentz invariance we need only two coupling constants, the values of the critical exponents are better known. First, in a perturbative calculation in powers of coupling constants, the exponents at critical points $A$ and $D$ will be identical. We may thus expect that in a lattice theory with short-range repulsion the transition is either into the time-reversal symmetry or chiral symmetry broken insulator, in either case with \cite{wetterich}
  \begin{equation}
  \nu=0.74 - 0.93
  \end{equation}
  \begin{equation}
  \eta_\Psi = 0.071 - 0.105.
  \end{equation}
  Note that since there is only a single Dirac field involved the numerical values of the exponents $\nu$ and $\eta_\Psi$ differ significantly from the large-N values of unity and zero, respectively. This raises hope that this non-trivial critical behavior may be observable in numerical simulations of lattice models.

 Although not of immediate relevance to graphene, it would still be of interest to determine the critical exponents at the chirally symmetric critical point C, which we proposed to control the critical behavior of the Thirring and the NJL model. We, however, are not aware of any analytical nor numerical study of the NJL model that goes beyond the leading order in large-N calculation. \cite{comment}

 \section{Fermi velocity and residue of quasiparticle pole}

   The critical exponents, as usual, govern for example the critical behavior of the gap on the insulating side of the transition, as mentioned in the introduction. In the present case, however, there are massless fermionic excitations on the metallic side, and one may wonder if and how the approach to the critical point is reflected onto these. Let us therefore generalize slightly and provide the support for the results already announced in the introduction.

     First, the usual scaling \cite{book} implies that at the cutoff $\Lambda/b$ electron's two-point correlation function near the critical point and at zero temperature satisfies
     \begin{equation}
     G= b^x \tilde{F} (b k,  b^z \omega, t b^{1/\nu}),
     \end{equation}
     where $\omega $ is the Matsubara frequency and $t\sim (V_c - V) >0$ is the transition's tuning parameter. Setting $t b^{1/\nu} =1$ we thus find the usual scaling law
   \begin{equation}
   G= t^{-x \nu} F(t^{-\nu} k, t^{-z \nu} \omega)
   \end{equation}
   where $F(x_1,x_2) = \tilde{F} (x_1,x_2,1)$ is a universal scaling function. From here we can extract the scaling of the Fermi velocity and quasiparticle residue as follows. First, if upon the analytical continuation to real frequencies $G$ has a pole at
   \begin{equation}
   \omega = v_F (t) k
   \end{equation}
   the scaling relation immediately dictates that
   \begin{equation}
   v_{F} (t) = t^{\nu (z-1)} v_F.
   \end{equation}

     Let us next set $\omega = 0$, take $k>0$, and let $t\rightarrow 0$. In this limit
     \begin{equation}
     G\sim \frac{1}{k^{1-\eta_k }}
     \end{equation}
     and therefore $F(x_1\rightarrow \infty, 0)\sim 1/x_1 ^{1-\eta_k}$. In order to cancel the $t$-dependence of the prefactor in Eq. (85) in this limit it must be that
     \begin{equation}
     x=1-\eta_k,
     \end{equation}
     where $\eta_k$ is the (momentum) anomalous dimension. Analogously, assuming that for $k=0$, $G\sim 1/\omega^{1-\eta_\omega}$, one finds that also
     \begin{equation}
     x= z (1-\eta_\omega).
     \end{equation}

     On the other hand, in the opposite limit $t>0$ and $\omega\rightarrow 0$, in the metallic phase at low energies we have fermionic quasiparticles. This implies that, for example, $F(0, x_2 \rightarrow 0)\sim 1/x_2 $, i. e.
     \begin{equation}
     G = \frac{Z }{ \omega},
     \end{equation}
     with $Z \sim t^{ (z-x) \nu}$.  Combining with the previous relation, the quasiparticle pole's residue behaves as
     \begin{equation}
     Z\sim t^{z \eta_\omega \nu}.
     \end{equation}
     For $z=1$, the two anomalous dimensions are the same, $\eta_k=\eta_\omega =\eta_\Psi$, where $\eta_\Psi $ is the Dirac fermion's anomalous dimension, and the scaling announced in the introduction follows. The special form of this relation for large number of Dirac components when besides $z=1$, it is also $\nu=1$, was previously proposed. \cite{herbut1} The quasiparticle residue vanishes upon the approach to the metal-insulator transition, as proposed long-ago by Brinkman and Rice, but here in a decidedly non-mean-field fashion.

 \section{Discussion}

   There are at least two obvious generalizations important for real graphene: the addition of spin, and the inclusion of the long-range tail of Coulomb repulsion. Adding spin would simply double the number of couplings for each symmetry, since each independent quartic term would then require a separate coupling in the singlet and in the triplet channels. The minimal internally consistent low-energy theory would then be the generalization of the Lorentz invariant Lagrangian in Eq. (70), with $g_\alpha=0$:
   \begin{equation}
   L_{int, lor} ^{spin} = \sum  g_{M,i}
   (\bar{\Psi}_{\alpha} M \sigma_{\alpha\beta}^i \Psi_{\beta} ) (\bar{\Psi}_{\gamma} M \sigma_{\gamma \delta}^i \Psi_{\delta} ),
   \end{equation}
 where the sum goes over $M=I, \gamma_{35}$, and $i=0, x, y, z$, with $\sigma_0= I_2$, and $g_{M,x}=g_{M,y}=g_{M,z}$.
 The Lagrangian with  $g_{\gamma_{35}, i}=0$ would represent the extended Hubbard model with on-site and nearest neighbor repulsion, considered before in the limit of large number of Dirac fermions in ref. 2. The interplay between the various instabilities in the theory equivalent to the above Lagrangian was recently studied in \cite{raghu}, where it was pointed out that the second-nearest-neighbor repulsion implies a negative coupling $g_{\gamma_{35}, 0}$ for example. The form of the above minimal spinful Lagrangian facilitates a systematic study of the metal-insulator transition in the Hubbard model, which will be a subject of a separate publication.

   Few comments on the importance of long-range tail of Coulomb interaction are also in order. Weak Coulomb ($\sim e^2 /r$) interaction is an (marginally) irrelevant perturbation at the Gaussian fixed point, and this remains true at the metal-insulator critical point at large-N  as well \cite{gonzales, herbut1, vafek}. Furthermore, the entire $\beta$-function for the charge coupling $e^2$ can be computed at large-N, and it does not exhibit any non-trivial zeroes. \cite{aleiner, son} On the other hand, several calculations show that by increasing the coupling $e^2$ beyond certain point and for small enough $N$ the system can be tuned through a metal-insulator transition  at which the chiral symmetry becomes spontaneously broken.\cite{gorbar, khveshchenko, leal, drut, stroutos}  The nature of such a putative metal-insulator transition is in our mind an open question at the moment. Whereas it is possible that it is described by new "charged" critical point \cite{case} corresponding to the non-trivial zero of the $\beta(e^2)$, it also seems conceivable that the charge is always irrelevant and that the transition is still in the universality class of the critical point C, in our nomenclature. Yet another possibility is a discontinuous transition. More work is obviously needed in order to be able to address this issue more conclusively. It may also be interesting to note that in the related bosonic problem, when a systematic expansion near four dimension is readily available, there are no charged critical points  in the theory to the leading order. \cite{remarkbook} This may also be contrasted with the well-known example of (albeit Lorentz invariant) scalar Higgs electrodynamics, for which the critical points, when they exist, are always charged. \cite{book, herb-tes}

     Probably the central message of this work is that provided Lorentz invariance becomes emergent near criticality, for the $p_z$-orbitals well localized on carbon atoms the Lagrangian may be taken to contain only two (or with the physical spin, four) coupling constants. If there are no intervening first-order transitions, one can  infer then that there are two possible continuous metal-insulator transitions, both governed by the same Gross-Neveu model in 2+1 dimensions, into the states that break either time-reversal or chiral symmetry. The residue of the quasiparticle pole on the metallic side plays the role of the metal's order parameter, and it vanishes continuously with a small critical exponent proportional to the fermion's anomalous dimension. In contrast, the specific heat coefficient \begin{equation}
     \lim_{ T\rightarrow 0} \frac{C_v}{T^2},
     \end{equation}
     being dependent on the Fermi velocity only, at the transition vanishes discontinuously from a finite value on the metallic side. Near the critical point and for the temperatures much below the bandwidth we may assume that the specific heat obeys  the scaling relation
     \begin{equation}
     C_v = T^{2/z} v_F ^{-2} R(\frac{T}{t^{z\nu}}).
     \end{equation}
     For small arguments the universal scaling function $R(x)$ behaves as $R(x)\sim x^{2(z-1)/z}$, so that in the metallic phase one finds the usual quadratic temperature dependence
     \begin{equation}
     C_v \sim T^2 v_F^{-2} t^{2\nu (1-z)}.
     \end{equation}
     Recognizing the proportionality of the specific heat coefficient as  $(v_F(t)) ^{-2}$ gives us yet another way to deduce Eq. (87). At the criticality, on the other hand,
     \begin{equation}
     \lim_{ T\rightarrow 0} \frac{C_v v_F ^2}{ T^{2/z} } = R(\infty),
     \end{equation}
     with $R(\infty)$ expected to be finite. When $z=1$, the specific heat coefficient near criticality jumps therefore from $R(0)$ in the metallic phase to $R(\infty)$ at the critical point, and finally to zero in the insulating phase.\cite{remarkgoldstone}

     Similarly, the optical conductivity near the metal-insulator transition will obey the scaling relation for $t>0$
     \begin{equation}
     \sigma(\omega) = H(\frac{\omega}{t^{z\nu}}) \frac{e^2}{h},
     \end{equation}
     with $H(x)$ as a universal function, and with
     \begin{equation}
     H(0)= \frac{\pi}{4},
     \end{equation}
     as the familiar universal dc conductivity per Dirac field in the metallic phase. \cite{herbut4} In contrast to the specific heat, there is no (non-universal) dimensionful quantity such as $v_F$ in the scaling expression for conductivity, and consequently $\sigma(0)$ is constant and universal in the entire metallic phase.  Right at the transition then
     \begin{equation}
     \sigma(\omega) = H(\infty) \frac{e^2}{h},
     \end{equation}
     so the  dc conductivity, while still universal, at the criticality should be different than in the metallic phase. Finally, in the insulator the dc conductivity vanishes, so that the dc conductivity, similar to the specific heat coefficient, in principle should show two universal discontinuities at the metal-insulator transition.

       One obstacle to experimental observation of these predictions is that it has not been possible yet to tune the parameter $\sim$(interaction/bandwidth) in graphene,  and sample different phases of the system. The application of the magnetic field, however, changes this, since the kinetic energy becomes completely quenched, and infinitesimal interaction immediately induces a finite gap. If the parameters of the system place it not too far from the metal-insulator transition, the gap $m$ would obey \cite{herbut2}
       \begin{equation}
       \frac{m}{v_F \Lambda}= (\frac{a}{l})^{z} G(l t^{\nu}),
       \end{equation}
       where $l$ is the magnetic length, $a=1/\Lambda$ is the lattice constant, and $t$ is the tuning parameter. $G(x)$ is a (universal) scaling function. The computation of the scaling function in the  large-N limit, and the consequences of this scaling relation for experiment were discussed at length before. \cite{herbut2} Here we only wish to underline that the emergent Lorentz invariance of the metal-insulator critical point, via its consequence that $z=1$, implies precise proportionality between the interaction gap and the Landau level separation at the criticality,
       \begin{equation}
       m = v_{F} \sqrt{B} G(0),
       \end{equation}
       where $G(0)$ is a universal number. Such a square-root magnetic field dependence of the gap is well known to arise from the long-range tail of the Coulomb interactions, but the above derivation serves to show that its origin may in principle lie in purely short-range interactions as well.

 \section{Summary}

   We have presented the theory of electrons interacting via short-range interactions on honeycomb lattice, and in particular,  determined the number and types of independent quartic terms in the low-energy Lagrangian. Metal-insulator quantum critical points and the concomitant quantum critical behavior were discussed, with the particular attention paid to the consequences of the emergent Lorentz invariance. The minimal internally consistent local Lagrangian for spinless fermions is shown to contain only two Gross-Neveu-like quartic terms. Generalizations that would include long-range Coulomb interaction or spin of electrons were briefly considered. We also discussed the critical behavior of several key physical quantities on the metallic side of the transition, such as the Fermi velocity, the residue of the quasiparticle  pole, specific heat, and the frequency dependent conductivity.

\section{Acknowledgement}

 This work was supported by the NSERC of Canada.

\appendix

\section{Fierz identity}

  For completeness, we provide the derivation of the Fierz identity. Assume a basis $ \{ \Gamma^a, a=1,..16 \}$ in the space of four-dimensional matrices, and choose $(\Gamma^a) ^\dagger = \Gamma^a = (\Gamma^a) ^{-1}$. Then any Hermitean matrix $M$ can be written as
  \begin{equation}
  M= \frac{1}{4} (Tr M \Gamma^a) \Gamma^a,
  \end{equation}
  with the summation over repeated indices assumed. This can be rewritten as
  \begin{equation}
   4 \delta_{li} \delta_{mj}  M_{lm} = \Gamma^a _{ml} \Gamma^a _{ij} M_{lm},
   \end{equation}
   and therefore it follows that
  \begin{equation}
   \delta_{li} \delta_{mj} = \frac{1}{4} \Gamma^a _{ml} \Gamma^a _{ij}.
   \end{equation}
   Applying this identity to the product of two matrix elements then yields
   \begin{equation}
   M_{ij} N_{mn} = \frac{1}{16} (Tr M \Gamma^a N \Gamma^b) \Gamma^b _{in} \Gamma^a _{mj}.
   \end{equation}
   Finally, this leads to the expansion of a quartic term as
   \begin{eqnarray}
   (\bar{\Psi}(x) M \Psi(x))  (\bar{\Psi}(y) N \Psi(y) ) =  \\ \nonumber
    - \frac{1}{16} (Tr M \Gamma^a N \Gamma^b)
    (\bar{\Psi}(x) \Gamma^b \Psi(y))  (\bar{\Psi}(y) \Gamma^a \Psi(x))
    \end{eqnarray}
    which is used in the text for $x=y$. The minus sign in the last line derives from the Grassmann nature of the
    fermionic fields.

\section{Diadic form of the asymmetric matrix}

  Any real $N$-dimensional matrix $M$ can obviously be written as
  \begin{equation}
  M= \sum_{i=1}^N  M_i \otimes e_i ^\top
  \end{equation}
  where $M_i ^\top = (M_{1i}, M_{2i}, ...M_{Ni})$, and $(e_i)_j = \delta_{ij}$. In Dirac notation,
  \begin{equation}
  M= \sum_{i=1}^N |M_i\rangle\langle e_i|,
  \end{equation}
  and
  \begin{equation}
  M^\top = \sum_{i=1}^{N} | e_i \rangle \langle M_i |
  \end{equation}
  is the transposed matrix. There exists such  a representation of the matrix $M$ in any basis of vectors
  $|\tilde{e}_i \rangle$, as can be seen by multiplying $M$ from the right with $1= \sum_i |\tilde{e}_i \rangle\langle \tilde{e}_i | $.

     Let us now form a related symmetric matrix $S_M = M^\top M$. Being symmetric, it can be written in the usual spectral form
 \begin{equation}
 S_M = \sum_{i=1}^N \mu_i |\mu_i \rangle \langle \mu_i|
 \end{equation}
where $\langle\mu_i|\mu_j\rangle =\delta_{ij}$. We can now write, however, the matrix $M$ in the
 particular eigenbasis of the associated symmetric matrix \cite{musicki} $S_M$
 \begin{equation}
 M = \sum_{i=1}^N |K_i \rangle \langle \mu_i|.
 \end{equation}
 From the definition of $S$ and its spectral form we see that $\langle K_i | K_j \rangle = \mu_i \delta_{ij}$, and therefore
 \begin{equation}
 M= \sum_{i=1}^N \sqrt{\mu_i} | \nu_i \rangle \langle \mu_i |
 \end{equation}
 where $|K_i \rangle = \sqrt{\mu_i} |\nu_i \rangle$, and $\langle \nu_i |\nu_j \rangle = \delta_{ij}$. For a general asymmetric matrix, the basis $\mu$ and $\nu$ are different, and the last equation generalizes the more familiar form for a symmetric matrix, where they are the same.

 \section{Renormalization group under Fierz constraints}

Here we provide an alternative formulation of the renormalization group transformation in presence of constraints imposed by the Fierz identity.  Let us demonstrate this method on the simplest example of the maximally symmetric theory. Instead of choosing two independent couplings and using Fierz transformation at intermediate stages of the calculation to transform any other generated quartic terms back into the chosen ones, one may use the kernel of the Fierz matrix to write the Lagrangian in terms only of the physical couplings from the outset, as in Eq. (53). The advantage of doing this is that no other quartic term besides the ones corresponding to the physical couplings can ever get generated then by the renormalization transformation. The set of couplings $\lambda_1$ and $\lambda_2$ is therefore closed under renormalization. The computation to the quadratic order then yields
\begin{equation}
\frac{d\lambda_1}{d\ln b} = -\lambda_1 - 24 \lambda_1 ^2 -72 \lambda_1 \lambda_2,
\end{equation}
\begin{equation}
\frac{d\lambda_2}{d\ln b} = -\lambda_2 - 72 \lambda_2 ^2 - 4 \lambda_1 ^ 2.
\end{equation}

  The connection to Eqs. (63)-(64) in the text can be established as follows. Since the Fierz transformations in this case imply
  \begin{equation}
  \vec{V} _\mu ^2 = -3 S^2,
  \end{equation}
  \begin{equation}
  \vec{V}^2 = S_\mu ^2 - 2 S^2,
  \end{equation}
  the Lagrangian $L_{int,max}$ in Eq. (41) can obviously also be written as
  \begin{equation}
  L_{int,max} = ( g_{D2}  -3 g_{B2} -2 g_{C1}) S^2  + (g_{A1} - g_{C1}) S_{\mu} ^2
  \end{equation}
  In the text we therefore have simply named the entire first bracket $g_{D2}$, and the second $g_{A1}$.   But these can be recognized as particular linear combinations of the physical couplings $\lambda_1$ and $\lambda_2$:
  \begin{equation}
  g_{D2}  -3 g_{B2} -2 g_{C1} = -2 \lambda_1 -12 \lambda_2,
  \end{equation}
  \begin{equation}
  (g_{A1} - g_{C1})= -2 \lambda_1.
  \end{equation}
  Such a connection is of course completely general, and in particular may be established between the three chosen couplings in the Eq. (70) and the "physical couplings" determined by the vectors in Eqs. (58)-(60).


\begin{thebibliography}{99}
\bibitem{rmp} For reviews, see V.P. Gusynin, S.G. Sharapov, J.P. Carbotte, Int. J. of Mod. Phys. B, {\bf 21}, 4611 (2007); A. H. Castro Neto, F. Guinea, N. M. R. Peres, K. S. Novoselov, A. K. Geim, Rev. Mod. Phys. {\bf 81}, 109 (2009).
\bibitem{herbut1} I. F. Herbut, Phys. Rev. Lett. {\bf 97}, 146401 (2006).
\bibitem{gonzales} J. Gonzalez, F. Guinea, M. A. H. Vozmediano,
Nucl. Phys. {\bf B424}, 595 (1994); Phys. Rev. B {\bf 59}, 2474 (1999).
\bibitem{vafek} O. Vafek, Phys. Rev. Lett. {\bf 98}, 216401 (2007).
\bibitem{herbutAF} I. F. Herbut, Phys. Rev. Lett. {\bf 99} 206404, (2007).
\bibitem{sorella} S. Sorella and E. Tosatti, Europhys. Lett. {\bf 19}, 699 (1992).
\bibitem{martelo}  L. M. Martelo, M. Dzierzawa, L. Siffert, and D. Baeriswyl, Z. Phys. B {\bf 103}, 335 (1997).
\bibitem{paiva} T. Paiva, R. T. Scalettar, W. Zheng, R. R. Singh, and J. Oitmaa, Phys. Rev. B {\bf 72}, 085123 (2005).
\bibitem{zhang1} Y. Zhang, Z. Jiang, J. P. Small, M. S. Purewal, Y.-W. Tan, M. Fazlollahi, J. D. Chudow, J. A. Jaszczak, H. L. Stormer, and P. Kim, Phys. Rev. Lett. {\bf 96}, 136806 (2006); Z. Jiang, Y. Zhang, H. L. Stormer, and P. Kim, Phys. Rev. Lett. {\bf  99}, 106802 (2007).
\bibitem{ong} Joseph G. Checkelsky, Lu Li, N. P. Ong, Phys. Rev. Lett. {\bf 100}, 206801 (2008); preprint arXiv:0808.0906.
\bibitem{khveshchenko} D. V. Khveshchenko, Phys. Rev. Lett. {\bf 87}, 206401 (2001); ibid. {\bf 87}, 246802 (2001).
\bibitem{gusynin} V. P. Gusynin, V. A. Miransky, S. G. Sharapov, and I. A. Shovkovy, Phys. Rev. B {\bf 74},
195429 (2006); E. V. Gorbar, V. P. Gusynin, V. A. Miransky, preprint arXiv:0710.3527.
\bibitem{herbut2} I. F. Herbut, Phys. Rev. B {\bf 75}, 165411 (2007); ibid. {\bf 76}, 085432 (2007); I. F. Herbut and B. Roy, {\bf 77}, 245438 (2008).
    \bibitem{alicea} J. Alicea and M. P. A. Fisher, Phys. Rev. B {\bf 74}, 075422 (2006); R. L. Doretto and C. Morais-Smith, Phys. Rev. B {\bf 76}, 195431 (2007).
\bibitem{geim} For a review, see A. K. Geim and K. S. Novoselov, Nature Materials {\bf 6}, 183 (2007).
\bibitem{herbut4} I. F. Herbut, V. Juri\v ci\' c, and O. Vafek, Phys. Rev. Lett. {\bf 100} 046403  (2008); I. F. Herbut, V. Juri\v ci\' c, O. Vafek, M. J. Case, preprint, arxive:0809.0725.
\bibitem{sachdev} L. Fritz, J. Schmalian, M. Mueller, and S.  Sachdev, Phys. Rev. B {\bf 78}, 085416 (2008).
\bibitem{remark1} By this definition the interaction that at large particle separation decays as a power law $V(r) \sim 1/r^a $ is short ranged provided that $a>2$.
    \bibitem{takahashi} See for example, Y. Takahashi, in {\sl Progress in Quantum Field Theory}, ed. by H. Ezawa and S. Kamefuchi (North Holland, 1986).
\bibitem{dwave} For the closely related chiral symmetry of d-wave superconductors, see I. F. Herbut,  Phys. Rev. B {\bf 66}, 094504 (2002); Phys. Rev. Lett. {\bf 88}, 047006 (2002); Phys. Rev. Lett. {\bf 94}, 237001 (2005); D. J. Lee and I. F. Herbut, Phys. Rev. B {\bf 66}, 094512 (2002); B. Seradjeh and I. F. Herbut, Phys. Rev. B {\bf 66}, 184507 (2002);  Z. Te\v sanovi\' c, O. Vafek, and M. Franz, Phys. Rev. B {\bf 65},  180511 (2002); M. Franz, T. Pereg-Barnea, D.E. Sheehy, and Z. Te\v sanovi\' c,  Phys. Rev. B {\bf 68}, 024508 (2003); I. O. Thomas and S. Hands, Phys. Rev. B {\bf 75}, 134516 (2007).
\bibitem{remark2} At the critical point at a finite Coulomb interaction in two-dimensions $z$ is always unity; see I. F. Herbut, Phys. Rev. Lett. {\bf 87}, 137004 (2001).
\bibitem{haldane} F. D. M. Haldane, Phys. Rev. Lett. {\bf  61}, 2015 (1988).
\bibitem{semenoff} G. W. Semenoff, Phys. Rev. Lett. {\bf 53}, 2449 (1984).
\bibitem{hou} C-Y. Hou, C. Chamon, and C. Mudry, Phys. Rev. Lett. {\bf 98}, 186809 (2007).
\bibitem{book} I. Herbut, {\sl A Modern Approach to Critical Phenomena}, (Cambridge University Press, Cambridge, 2007).
\bibitem{brinkman} W. F. Brinkman and T. M. Rice, Phys. Rev. B {\bf 2}, 4302 (1970).
\bibitem{kohmoto} Y. Hasegawa, R. Konno, H. Nakano, M. Kohmoto, Phys. Rev. B {\bf 74}, 033413 (2006).
    \bibitem{herbutT} I. F. Herbut, Phys. Rev. B. {\bf 78}, 205433 (2008).
\bibitem{musicki} Dj. Mu\v sicki and B. Mili\' c, {\sl Matematicke osnove teorijske fizike}, (Univerzitet u Beogradu, Beograd, 1984), sec. 5.3 (in Serbian).
\bibitem{remarkshell} Here we treat the frequency integrals in the loops analogously to momentum integrals. For an alternative, see ref. 2, for example.
\bibitem{kaveh} K. Kaveh and I. F. Herbut, Phys. Rev. B {\bf 71}, 184519 (2005).
\bibitem{vasilev} A. M. Vasil'ev, S. E. Derkachov, N. A. Kilev,
and A. S. Stepanenko, Teor. Mat. Fiz. {\bf 92}, 486 (1992); {\sl
ibid.} {\bf 97}, 364 (1993);
\bibitem{gracey} J. A. Gracey, Int. J. Mod. Phys. A {\bf 9}, 727 (1994).
\bibitem{karkkainen} L. K\"{a}rkk\"{a}inen, L. Lacaze, P. Lacock, and B.
Petersson, Nucl. Phys. B {\bf 415}, 781 (1994); {\bf 438} 650(E) (1995).
\bibitem{wetterich} L. Rosa, P. Vitale, and C. Wetterich, Phys. Rev.
Lett. {\bf 86}, 958 (2001); F. H\"{o}fling, C. Novak, and C. Wetterich, Phys. Rev. B {\bf 66}, 205111 (2002).
\bibitem{christofi} S. Christofi, S. Hands, C. Strouthos, Phys.Rev. D {\bf 75}, 101701 (2007), and references therein.
\bibitem{nambu} Y. Nambu and G. Jona-Lasinio, Phys. Rev. {\bf 122}, 345 (1961).
\bibitem{calabrese} In the purely bosonic $\Phi^4$-theories, the $O(3)$ symmetry is currently believed not to emerge at the critical point out of $Z_2 \times O(2)$, in three dimensions. The critical behavior is governed by the "biconal" fixed point, which however, appears to be extremely close the $O(3)$-symmetric point in the coupling space. See, P. Calabrese, A. Pelissetto, E. Vicari, Phys. Rev. B {\bf 67}, 054505 (2003).
\bibitem{remarkatomic} The terms allowed in the atomic limit are already contained in the terms $S^2$ and $V_1 ^2$.
\bibitem{comment} See, nevertheless, \cite{christofi} for numerical results on Thirring models with more than one Dirac fermion.
\bibitem{raghu} S. Raghu, Xiao-Liang Qi, C. Honerkamp, S.-C. Zhang, Phys. Rev. Lett. {\bf 100}, 156401 (2008).
\bibitem{aleiner} I. L. Aleiner, D. E. Kharzaev, and A. M. Tsvelik, Phys. Rev. B {\bf  76}, 195415 (2007)
\bibitem{son} D. T. Son, Phys. Rev. B {\bf 75}, 235423 (2007); J. E. Drut and D. T. Son,  Phys. Rev. B {\bf 77}, 075115 (2008).
\bibitem{gorbar} E. V. Gorbar, V. P. Gusynin, V. A. Miransky, and I. A. Shovkovy , Phys. Rev. B {\bf 66},  045108, (2002).
\bibitem{leal} D. V. Khveshchenko and H. Leal, Nucl. Phys. B {\bf 687}, 323 (2004).
\bibitem{drut} J. E. Drut and T. A. Lahde, Phys. Rev. Lett, {\bf 102}, 026802  (2009).
\bibitem{stroutos} S. Hands and C. Strouthos, Phys. Rev. B {\bf 78} 165423, 2008.
\bibitem{case} O. Vafek and M. J. Case, Phys. Rev. B {\bf 77}, 033410 (2008).
\bibitem{remarkbook} See the ref. 21, and the problem 8.10 in ref. 25.
\bibitem{herb-tes} I. F. Herbut and Z. Te\v sanovi\' c, Phys. Rev. Lett. {\bf 76}, 4588 (1996); ibid. {\bf 78}, 980 (1997); I. F. Herbut, J. Phys. A: Math. Gen. {\bf 30}, 423 (1997).
\bibitem{remarkgoldstone} For spinless fermions considered here, both insulators break the discrete Ising symmetry, so there are no Nambu-Goldstone bosons.

\end{thebibliography}
\end{document}